\documentclass[a4paper,11pt]{article}
\usepackage{pos}
\usepackage[normalem]{ulem}

\def\mnras{MNRAS}

\def\apj{ApJ}
\def\aapr{A\&ARv}
\def\aj{AJ}

\title{The Readiness of EVN Telescopes for the SKA-VLBI Era}

\author*[a,b,c]{Mar\'{\i}a J. {Rioja}}
\author[a]{Richard {Dodson}}

\affiliation[a]{ICRAR, M468, The University of Western Australia, 35 Stirling Hwy, Crawley, Western Australia, 6009}

\affiliation[b]{CSIRO Astronomy and Space Science, PO Box 1130, Bentley WA 6102, Australia}

\affiliation[c]{Observatorio Astron\'omico Nacional (IGN), Alfonso XII, 3 y 5, 28014 Madrid, Spain}

\emailAdd{maria.rioja@icrar.org}
\emailAdd{richard.dodson@icrar.org}

\abstract{
The application of VLBI to scientific problems has undergone a relentless expansion since its conception, yet the potential for further expansion is still large.
We are on the cusp of revolutionary progress given the arrival of a host of next-generation instruments.
Over the last few years the community has been working hard to ensure the SKA
design includes the capability to enable multiple simultaneous
tied-array beams, which is a crucial technology to deliver
ultra-precise astrometry and improve survey speed capabilities.
However, to reach the full potential requires that the network of antennas is upgraded to match the SKA capabilities.
We identify multiple-pixel technology, on large telescopes and connected arrays, as a crucial missing component and here will make recommendations for the upgrade path of the partner EVN (and other network) telescopes.
Our feasibility studies on SKA-VLBI suggest an order of magnitude improvement in the precision and also in the frequency range at which astrometry can be performed today, if the full network has the required capabilities. 
}

\FullConference{%
15th European VLBI Network Mini-Symposium and Users Meeting\\
11-15 July, 2022\\
University College Cork, Ireland}

\begin{document}
\maketitle

\section{Introduction}

Precision astrometry measurements add a new dimension to the research of many astrophysical fields. It
has provided deep insight into the astrophysical processes in a huge range of environments, and is a probe for fundamental properties of the Universe. 
VLBI has traditionally provided the highest astrometry precision measurements and the next-generation of instruments 
hold the potential for an order of magnitude of improvement. 
However, the current conventional 
astrometric methods are limited by systematics in most cases. Hence we need a 
commensurate improvement in methods as well as in sensitivity for the next-generation of instruments.

Since the beginnings of VLBI, there has been a relentless quest for increase\typeout{d} of { astrometric} accuracy and wider applicability.
In the last decade there have been huge strides in achieving the full astrometric potential, arising
from advanced phase referencing (PR) calibration strategies \citep[i.e. GeoBlocks:][]{reid_04} 
that accurately  compensate for the dominant tropospheric
propagation residual errors that have led to a few tens of micro-arcsecond precision measurements,
most notably at $\sim$22\,GHz. 

Instead the field of astrometry at lower frequencies ($<$8\,GHz) has lagged behind, 
with an order of magnitude larger errors at L-band ($\sim$1.6\,GHz). This regime is dominated
by residual ionospheric propagation effects that pose a rather different  set of challenges. Chief of those challenges is the fact that the ionospheric effects have a strong spatial structure (i.e. they are direction-dependent), which limits the use of observations of a reference source (necessarily along a different line of sight than that of the target) to correct for the atmospheric errors. 
We note as an aside the multi-frequency advanced PR calibration strategy \cite{dodson_17}, where ionospheric residuals along the line of sight from the target are removed with ``ICE-blocks'' without using a reference source. 
Thus the direction-dependent corrections are not required.

The arrival of the SKA, currently under construction, will focus on the lower frequencies and will revitalise all aspects of VLBI astronomy at these wavelengths with joint observations with EVN telescopes.
Among these, the ultra precise astrometric capability is of great importance and is a scientifically driven motivation in the SKA-VLBI era; the goal is to
improve by an order of magnitude both the astrometric accuracy and the range of applicable frequencies \citep{ska_vlbi, jj_wp10}.
The high sensitivity and long baselines of SKA-VLBI observations will result in a much reduced thermal noise level and high spatial resolution.  Therefore this goal is achievable, as long as a
sufficiently accurate ionospheric phase-calibration strategy is in place.  

This paper discusses the way to achieve this using a next-generation method, namely MultiView \citep{rioja_17}, and an upgrade of the network  telescopes to implement its requirements for optimal performance.
Section 2 describes the basics of MultiView, presents current empirical demonstrations with existing instruments and our estimates for the expected performance in the SKA-VLBI era. 
Section 3 describes the technological developments relevant to astrometry for the telescope network, namely \typeout{multiple beam} multiple-pixel capabilities.  Section 4 are the conclusions.

\section{cm/m-VLBI Microarcsecond astrometry using MultiView methods}

MultiView \citep{rioja_17} is a next-generation calibration method that has the potential for optimum correction of the dominant ionospheric residual errors, which are the main challenge and limit the measurements at the SKA frequency range with PR.
The result is ultra high precision astrometry measurements with wide applicability, that is, to many sources and across a much wider frequency range than  can be used today \citep{rioja_20}.

\begin{figure} 
    \centering
    \includegraphics[width=0.75\textwidth]{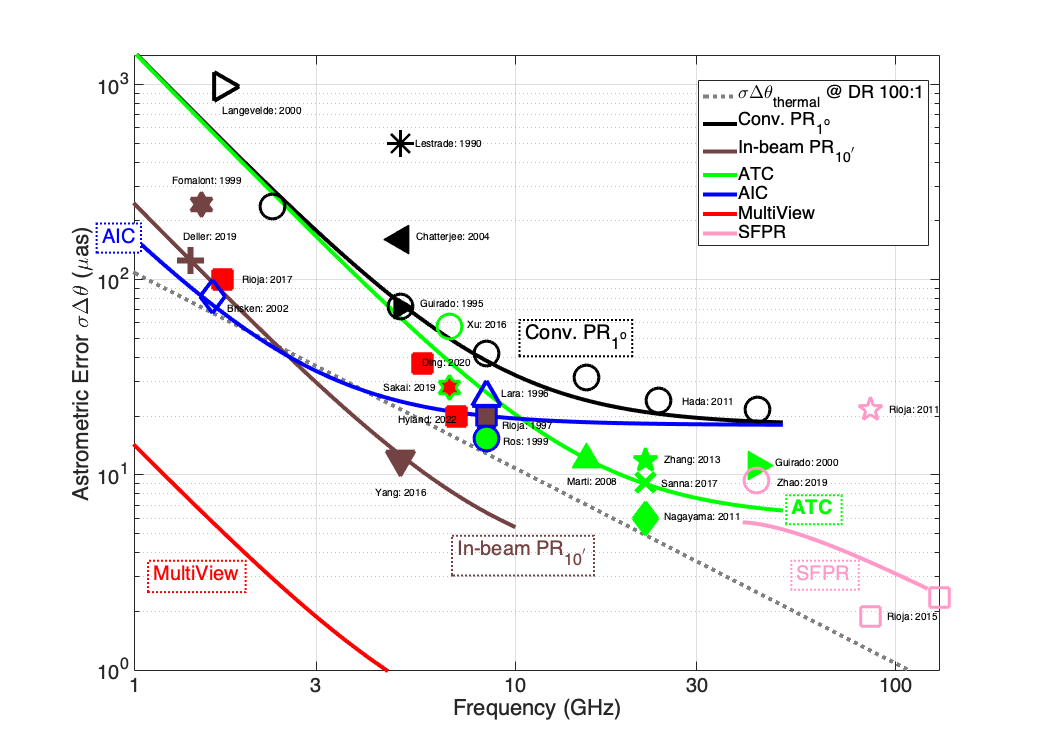}
    \caption{
    Illustration of theoretical systematic limits for a range of astrometrical techniques (solid lines: for conventional PR in black and brown for in-beam; Advanced Tropospheric Calibration (ATC) in green; Advanced Ionospheric Calibration (AIC) in blue) compared to actual observational results from the literature shown with symbols, taken from \citet{rioja_20}. This emphasises how these limits are greater than the thermal limit for a dynamic range 100:1 (dotted line) and that the systematics dominate observationally except for the next-generation methods, MultiView (red) and SFPR (pink). See \citet{rioja_20} for details.
    \label{fig:now}}
\end{figure}

Standard PR strategies using a single calibrator result in uncorrected systematic residual phase errors caused by   the 
spatial structure in the atmospheric propagation effects (even when the sources are close i.e. $\sim1^o$).
These systematics strongly 
affect 
the analysis and 
impose limits in the astrometric accuracy. Figure \ref{fig:now} illustrates these astrometric limits across the frequency range.
Instead, MultiView calibration uses 
observations of multiple calibrators surrounding the target 
and combines their phases with spatial
interpolation to effectively result in using a virtual calibrator at $\sim0^o$ angular separation.
PR performance increasingly deteriorates towards the low frequency regime, with errors much larger than at the higher frequencies, such as 22 GHz, limiting its application. 
A clear example of the residual ionospheric propagation effects becoming the dominant source of errors at observing frequencies less than $\sim\,8$ GHz is to be found in the analysis of 6.7 GHz methanol maser observations in the BeSSeL project, using similar PR with GeoBlocks strategies than for 22 GHz water maser observations. 
The precision of the results from 2016 \citep{xu_16}, 
are significantly worse 
than the re-analysis \citep{wu_19} with a variation on the MultiView strategy \citep{reid_17}.

\begin{figure}
    \centering
    \includegraphics[width=0.75\textwidth]{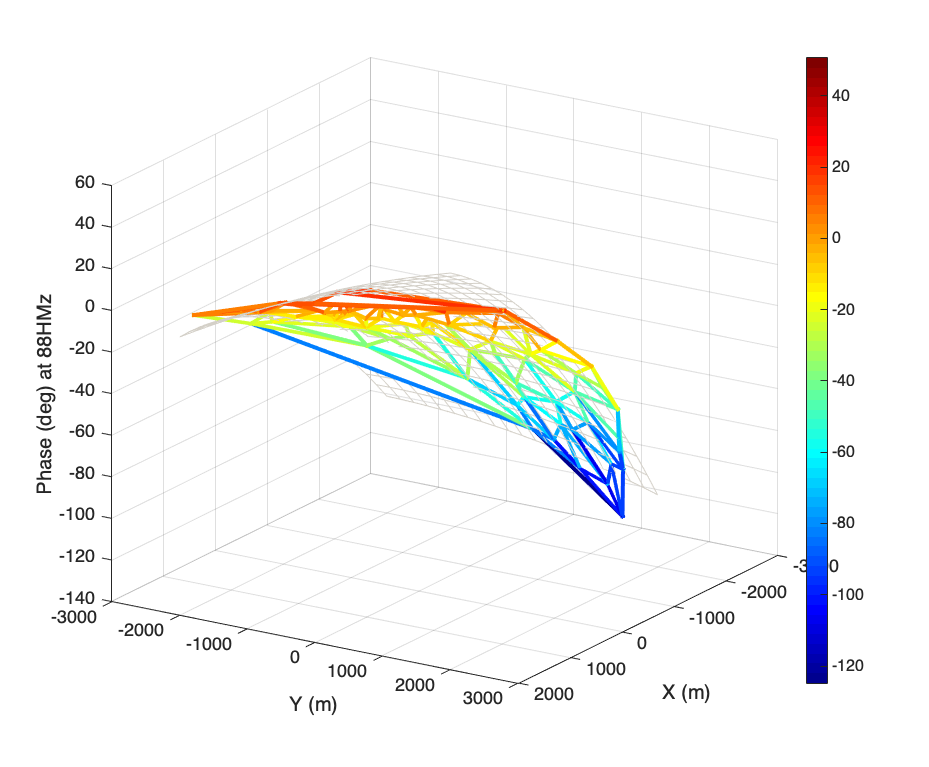}
    \caption{Example of an ionospheric residual phase screen as observed over the MWA at 88MHz after direction independent and bandpass calibration.
    Shown as a mesh are the station-based calibration phases above the site for one direction, color-coded in degrees, with the X and Y axis in meters for the 128 stations (at the nodes of the mesh).     
    The wireframe shows the second order fit to the data, underlining the high curvature of the surface. The typical residuals to a planar fit are about 1mTECU.
    See \citet{rioja_22} for details.}
    \label{fig:phase_sc}
\end{figure}

MultiView calibrators can be further away than for PR  because the solutions are interpolated rather than transferred.
This has been demonstrated at 1.6\,GHz \citep[][]{rioja_17}, 8\,GHz \citep[][]{ding_20_mv,hyland_22} and 6.7\,GHz \citep[][]{hyland_22b} with calibrators up to 6$^o$ from the target.
Encouraged by the outstanding performance, explorations in PR corner cases are under way, with observations at very low elevations, very low (0.3GHz) and high (43GHz) frequencies, where we are exploring the application for next-generation instruments such as SKA, ngVLA-LONG  and FAST.
For maximum precision one requires simultaneous observations of (nearby) calibrators uniformly distributed surrounding the target.
Nominally, following from empirical ionospheric spatial structure studies with MWA \cite{rioja_22}, the expectations of residual errors with MultiView would be about 1 mTECU.
See Figure \ref{fig:phase_sc} for an example of the ionospheric phase screens as observed with the MWA. 
Residuals of this level would result in MultiView systematic astrometric errors at the 1 micro-as level above $\sim$5GHz (see Table \ref{tab:inbeam_mv}) with the final error to be comprising of the additional contributions set by the thermal noise  or measurement errors, the dynamic range of the image and the stability of the reference points selected within the sources for multi-epoch comparisons.  This precision is more than an order of magnitude improvement over the current limits. 

\begin{figure} 
    \centering
    \includegraphics[width=0.75\textwidth]{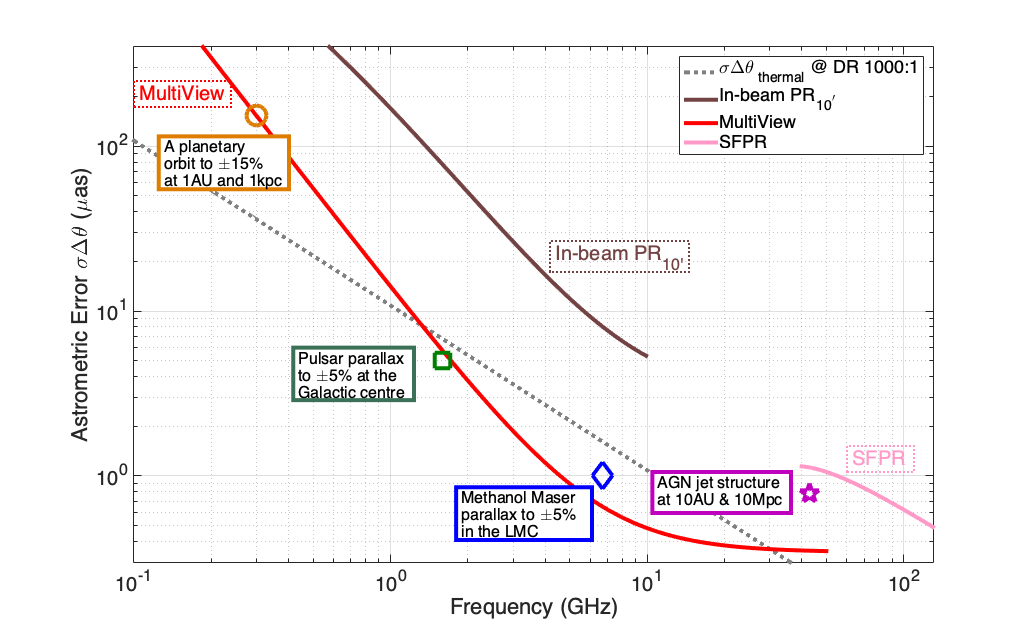}   
    \caption{
The theoretical thermal (dotted line) and systematic (solid lines) limits for the next-generation of instruments and methods, taken from \citet{rioja_20}. This illustrates the astrometric improvements of MultiView, which matches the thermal limit of a dynamic range 1000:1. Some suggested astrometric projects are indicated. See \citet{rioja_20} for details.    
\label{fig:tomorrow}}
\end{figure}

The improvement that MultiView can provide to VLBI astrometric observations has been demonstrated with an increasing number of empirical astrometric measurements, showing outstanding performance compared to standard PR methods, reaching the
thermal noise limit of current VLBI networks, as predicted by our error analysis.  
Thus we are confident in our estimates of an order of magnitude improvement for
SKA-VLBI, assuming an upgraded network of antennas that match the SKA
capabilities.  This is predicated on the sensitivity improvement from
the increased collecting area (and bandwidth) and the quasi-perfect compensation of
systematic atmospheric effects, as provided by MultiView (see Figure \ref{fig:tomorrow}). 
Note that MultiView is expected to achieve an order of magnitude improvement compared to in-beam PR.

As stated above, 
the ultimate limit, after correcting for the currently dominant atmospheric propagation errors, we expect to be related to intrinsic source structure effects and the definition and stability of the reference points 
over time.
To alleviate their impact, and other potential phase ambiguity related issues in the analysis, we plan for an over-determined fit to the phase plane, i.e. to use more than the minimum number of three calibrators surrounding the target, and a dense network (to provide uv-coverage) of moderate to large sized telescopes for joint MultiView observations. 

To provide simultaneous observations of all sources we identify multiple-pixel technology, 
on large telescopes and connected arrays, as the crucial missing component. In the next section we describe the technology and make recommendations for the upgrade path of the partner EVN (and other network) telescopes. 

\begin{table}
    \caption{
    Table to characterise the performance of MultiView and its feasibility for current and next-generation instruments, across the SKA spectrum, taken from \citet{rioja_20}.
    Col. 1 is the observing frequency, 
    Col. 2 is the estimated systematic astrometric error using MultiView, as discussed in \citet{rioja_20}.
    Col. 3 is the number of 100$\sigma$  calibrator sources for current arrays, expected within the primary beam of a single pixel 20 m antenna if FoV< $1^o$, otherwise 1$^o$ (marked with $^{\dagger}$), 
    calculated using the source count prediction from \citet{TREC}.
    Col. 4 is the same as Col. 3, but using the sensitivity for SKA-VLBI Phase-1 (in brackets for Phase-2 at the higher frequencies) \citep{jj_wp10} that are strong enough to exceed the MultiView systematic limits ($\sim\,1000\sigma$, see \citet{rioja_20} for details).
    We note that the number of in-beam sources for ngVLA-LONG observations would fall between SKA Phase 1 and 2 estimates.
    Based on Col. 3 \& 4, simultaneous (e.g. within primary beam, in-beam) MultiView would be feasible at frequencies $<$1.4, $<$2 and $<$6.7\,GHz, with the sensitivities of current VLBI, SKA-VLBI Phase 1 and Phase 2, respectively. At higher frequencies, MultiView is possible using nodding observations or using simultaneous observations with sites with multi-antennas (e.g. ngVLA-LONG) or subarraying (e.g. SKA) capabilities. 
    \label{tab:inbeam_mv}}
    \centering
    \begin{tabular}{|c|c||l|l|}
    \hline
    Frequency & MultiView error & No. in-beam &No. in-beam \\
    $\nu$ & $\sigma\Delta\theta^{MV}$ & sources for  & sources for \\
     (GHz) & ($\mu$as) &  current arrays & SKA-VLBI \\
        \hline 
        0.3 & 150    & 1.2$^{\dagger}$ & 14$^{\dagger}$ \\
        0.9 & 17     &3.5 & 15  \\
        1.6 & 6     & 2.9 & 5.5  \\
        5.0 &$\sim$1& 0.4 & 0.4 (6) \\
        8.0 & $\sim$1& 0.1 & 0.1 (2) \\
        15.0& $\sim$1& 0.0 & 0.0 (0.4) \\
        \hline
    \end{tabular}
\end{table}

\section{
Implementation of MultiView Technological requirements for the telescope network}

\begin{figure}
    \centering
    \includegraphics[width=0.65\textwidth]{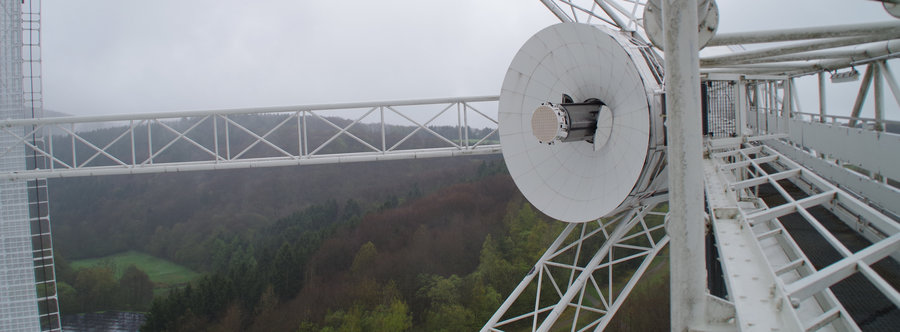}
    \includegraphics[width=0.3\textwidth]{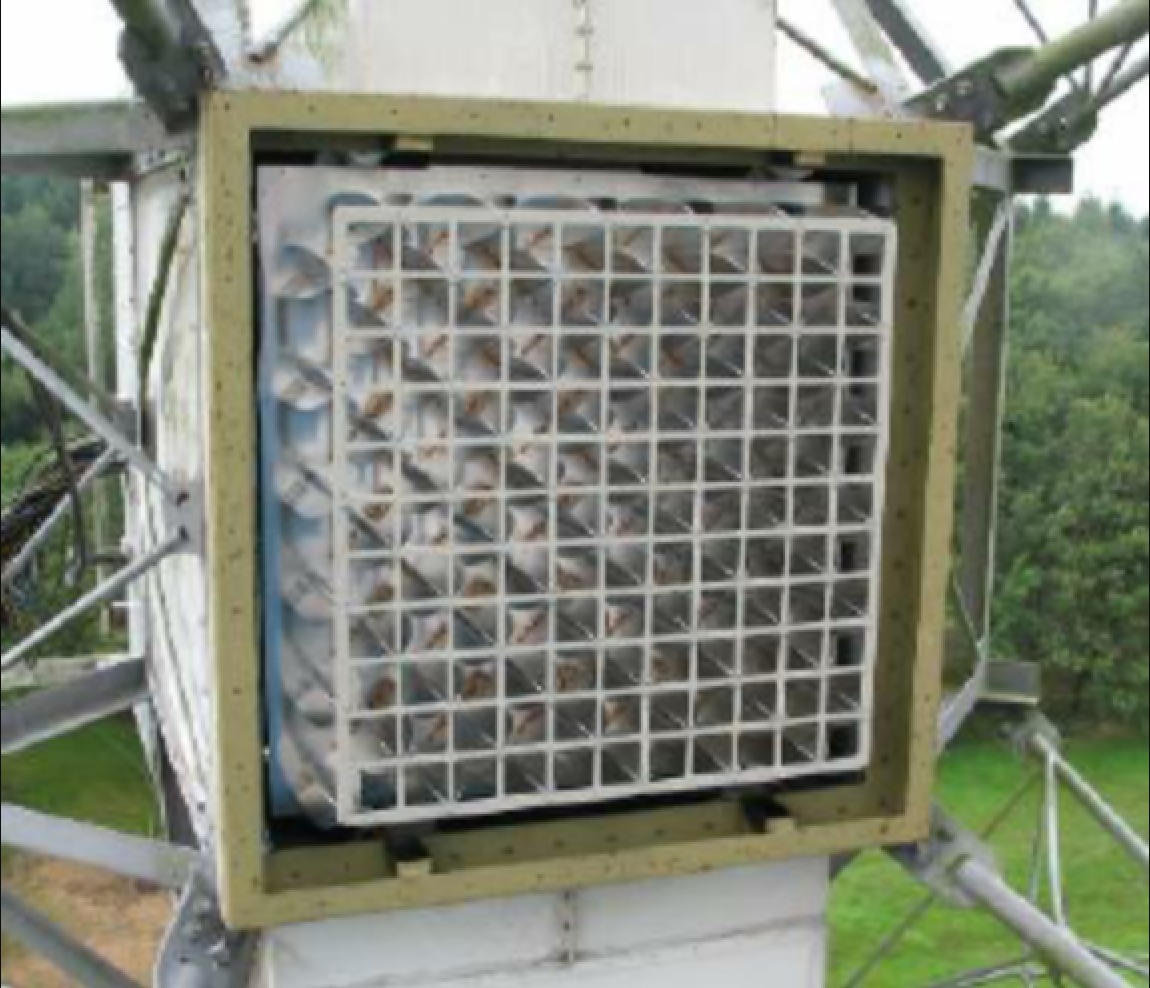}
    \caption{Left: The CSIRO MKII PAF installed at the Effelsberg telescope prime focus. 
    MPIfR are now developing their own version with increased capabilities. Of particular importance for multiple-pixel VLBI will be the ability for the PAF beams to track the same point in the sky during the parallactic angle rotation. 
    Right: The APERTIF PAF in WRST.
    }
    \label{fig:eff_paf}
\end{figure}
\typeout{\it Exec Summary: MultiView optimum performance comes from high SNR and simultaneous multi-source observations. Large collecting areas have small FoV. Multiple beam technologies bring together high SNR and large FoV. These provide the observational requirements for MultiView. 
SKA design requirement already in place. Time to act on telescope networks, plan of action: detailed specification of requirements and practical demonstrations.}


Optimum MultiView performance comes from simultaneous high-SNR observations of the multiple sources involved. This  translates to observational requirements, namely sensitive observations with large collecting areas and (relatively) wide FoVs. 
Multiple-pixel technology is fundamental to bring the two requirements together, for large telescopes and connected arrays across the network. 
These MultiView observational requirements have driven the
community efforts that resulted in an Engineering Change Proposal to ensure that the SKA design includes the innovative capability 
to enable multiple simultaneous tied-array beams from the connected  phased-up array and/or from subarraying, which has been approved.
These support a number of concurrent pencil beams ranging in number between 4 full-sensitivity 
VLBI beams with 2.6 GHz bandwidth and up to 46 beams in total, with bandwidth tradeoffs \citep[for details see][]{jj_wp10}. These beams can simultaneously point in any direction within the individual single-antenna FoV with the full sensitivity of the connected array, or have even wider separations by using subarraying. 
 
The remaining missing component is to upgrade the rest of the network of large telescopes and connected arrays, to match the capabilities of SKA. 
To deliver this upgrade path for the partner EVN (and other) telescopes 
we need to both define the technological multiple-pixel VLBI beam requirements and carry out practical end-to-end demonstrations to discover the key issues. 
These two activities necessarily depend on each other, as one sets what can be done and the other sets what must be done.
\typeout{\bf (are these written in reverse order???)}
The benefits of innovative multiple-pixel technologies available to increase the FoV such as Phased Array Feeds (PAFs) and Multi-Beam receivers are recognized within the EVN, with the largest telescopes already equipped or with plans to do so. 
For example, Effelsberg (100m, Germany) and Lovell (76m, UK) telescopes are equipped with CSIRO MKII PAFs, the Sardinia Radio Telescope (64m, Italy) has multi-beam receivers, and WSRT array is equipped with the APERTIF PAFs system.
A 100m telescope such as Effelsberg with a 25-beam PAF has the same FoV as a 20m and a connected array with multiple tied-array beams, such as WRST, has the same FoV as the individual elements. 
Nevertheless, its application to VLBI has not been implemented so far, and plans to do it have a very low priority.
Figure \ref{fig:eff_paf} shows the current PAFs on Effelsberg and WSRT.

Such an upgrade will benefit the operations of the EVN as an stand alone instrument  as well as providing 
an order of magnitude improvement in astrometric precision for SKA-VLBI observations. 
Other than SKA, the multiple-pixel \typeout{\bf technology/} capability is a part of the design of other next generation instruments, such as FAST and ngVLA-LONG.

\section{Conclusions}

The arrival of sensitive next generation instruments brings exciting opportunities and challenges for VLBI observations \citep{rioja_20}.
Among the former, the realization of ultra precise astrometric measurements that will enable the addressing of a host of innovative open questions in astrophysics. 
The list of challenges includes the readiness of the telescope network to reach the astrometric potential in joint observations with existing telescopes.
This paper is mainly concerned with an upgrade of the EVN telescopes, for improvement and benefits of joint observations with SKA and also as a stand-alone instrument. The proposed multiple-pixel capability for large telescopes and arrays is predicated on the  
next-generation calibration methods and their observational requirements.

MultiView was originally conceived to address the poor astrometric performance of conventional PR methods at low (<8 GHz) frequencies. 
MultiView has clearly demonstrated superior performance, with increased astrometric precision from removing the dominant 
ionospheric errors and wide applicability, to many sources, at
frequencies $<$8\,GHz using existing instruments. 
Based on this outstanding performance a number of efforts are ongoing to extend the MultiView method beyond its original scope, to higher and lower frequencies, as well as to better characterise the performance limits in corner cases. 
These include astrometric observations at very low  elevations and with ever wider angular separations. 
The on-going investigations at higher frequencies show very promising outcomes for the correction of the tropospheric propagation medium effects as well as the ionospheric effects. 


SKA has 
adopted the new technologies required for the next-generation of calibration method MultiView, i.e. multiple tied-array beam technologies in this particular case. 
It is imperative that we ensure that the keystone telescopes that make up the EVN  are equally prepared for
these new  techniques. 
Thus it is urgent that a roadmap for the telescope network, that includes fleshing out the requirements of individual EVN partners and for the various technological multiple-pixel solutions, is developed. As part of this, further practical end-to-end demonstrations are vital to define what is needed (e.g. number of beams, angular range) that will affect the technological options.
These technological upgrades will impact the VLBI observations of the EVN with the SKA, but also the capabilities of the  
EVN as a stand alone array both for astrometry and survey speed.
Furthermore, these considerations are also of interest to FAST-VLBI and ngVLA-LONG and space VLBI astrometry.

\end{document}